\documentclass[12pt,preprint]{aastex}

\newcommand{\exhydra}{EX~Hydrae}
\newcommand{\exhya}{EX~Hya}
\newcommand{\phispin}{\ifmmode{\phi_{67}}\else{$\phi_{67}$}\fi}
\newcommand{\phiorbit}{\ifmmode{\phi_{98}}\else{$\phi_{98}$}\fi}
\newcommand{\phiorbitmin}{\ifmmode{\phi_{98,{\rm 
min}}}\else{$\phi_{98,{\rm min}}$}\fi}
\newcommand{\phispinfold}{\ifmmode{\phi_{67,{\rm 
fold}}}\else{$\phi_{67,{\rm fold}}$}\fi}
\newcommand{\phiorbitfold}{\ifmmode{\phi_{98,{\rm 
fold}}}\else{$\phi_{98,{\rm fold}}$}\fi}
\newcommand{\kms}{\ifmmode{{\rm km~s}^{-1}}\else{km~s$^{-1}$}\fi}

\shorttitle{Accretion Column Structure of Cataclysmic Variables}
\shortauthors{Hoogerwerf et al.}

\begin{document}


\title{ACCRETION COLUMN STRUCTURE OF MAGNETIC CATACLYSMIC VARIABLES FROM X-RAY
SPECTROSCOPY}


\author{R.\ Hoogerwerf, N.~S.\ Brickhouse}
\affil{Smithsonian Astrophysical Observatory, Harvard-Smithsonian
Center for Astrophysics, 60 Garden Street, MS 31, Cambridge, MA 02138}
\email{rhoogerwerf@cfa.harvard.edu,nbrickhouse@cfa.harvard.edu}

\and

\author{C.~W.\ Mauche}
\affil{Lawrence Livermore National Laboratory, L-473, 7000 East Avenue,
Livermore, CA 94550}
\email{mauche@cygnus.llnl.gov}


\begin{abstract}
Using {\it Chandra}\/ HETG data we present light curves for individual
spectral lines of \ion{Mg}{11} and \ion{Mg}{12} for \exhydra, an
intermediate-polar type cataclysmic variable. The \ion{Mg}{11} light
curve, folded on the white dwarf spin period, shows two spikes that
are not seen in the \ion{Mg}{12} or broad-band light curves.
Occultation of the accretion column by the body of the white dwarf
would produce such spikes for an angle between the rotation axis and
the accretion columns of $\alpha = 18\degr$ and a height of the
\ion{Mg}{11} emission above the white dwarf surface of $\la0.0004$
white dwarf radii or $\la 4$~km. The absence of spikes in the
\ion{Mg}{12} and broad-band light curves could then be explained if
the bulk of its emission forms at much larger height, $>0.004$ white
dwarf radii or $>40$~km, above the white dwarf surface, although this
is not consistent with the predictions of the standard Aizu model of
the accretion column. 
\end{abstract}



\section{Introduction}
In magnetic cataclysmic variables (CVs) the primary is a highly
magnetized ($B \sim 0.1$--100 MG) white dwarf whose field controls the
accretion flow close to the white dwarf, leading to a shock and
accretion columns that radiate mainly in X-rays.  Unfortunately, the
size scales associated with these shocks and subsequent cooling flows
are small, on the order of the white dwarf radius or smaller and
hence impossible to resolve spatially with any instrumentation
available today. This is unfortunate, because measurements of the
density and temperature structure in accretion columns would provide
valuable data with which to test our understanding of accretion models
and the many physical processes involved. Through eclipse mapping
\citep{hor1985}, Doppler tomography \citep{mar1988}, Stokes imaging
\citep*{pot1998}, and Roche tomography \citep{dhi2001} the companion,
accretion disk/stream/curtain, and magnetic foot points have all been
mapped. A similar treatment of the accretion column has not been
possible until now.

\exhydra\ is the brightest intermediate polar (IP) type CV in the sky,
with a reliable inclination \citep*[$i =
77\fdg2\pm0\fdg6$,][]{beu2003,hoo2004} determined from the partial
eclipse in its light curve.  The X-ray light curves of \exhya, folded
on the white dwarf spin period, 68~min, show a sinusoidal modulation
of increasing amplitude with wavelength. Two models have been
suggested for this modulation. The first is the accretion curtain model
\citep*{ros1988}, in which the pre-shock material absorbs the emission
created in the post-shock region. Since the pre-shock material is
confined by the magnetic field, its projected column density changes
as the white dwarf rotates, hence changing the amount of absorption
\citep[see figures 7 and 8 of][]{ros1988}. The second is the
occultation model \citep*[e.g.,][]{all1998,muk1999}, in which the
modulation is due to occultation of the emission by the body of the
white dwarf.  Unfortunately, using broadband light curves, as has been
done in the past, it is impossible to distinguish between the two
models.

Using individual spectral-line light curves can break the degeneracy
between the accretion curtain model and the occultation model.  In
this letter, the X-ray light curve of individual spectral lines of
He-like \ion{Mg}{11} is used, in the context of the
occultation model, to determine the angle of the accretion column and
the height above the white dwarf surface of the emission region. With increased signal-to-noise
ratio (SNR), this new diagnostic technique could be applied to lines
that form at different temperatures and densities. Such an analysis
would enable a mapping of the distribution of line emissivities, and
hence the temperature and density structure, along the accretion
column.

\section{Observations, Reduction, and Light Curves}

\exhya\ was observed by {\it Chandra}\/ using the High Energy
Transmission Grating (HETG) in combination with the Advanced CCD
Imaging Spectrometer in its spectroscopy layout (ACIS-S) on 2000 May
18 for 60 ks (ObsID 1706). The observation was continuous and covers $\sim$10
orbital revolutions of the binary system and $\sim$15 white dwarf
revolutions. We reduced the data using the {\it Chandra} Interactive
Analysis of Observations (CIAO version 3.0) software
package\footnote{http://cxc.harvard.edu/ciao/}.

Light curves were constructed using the CIAO tool {\it dmextract}.  We
generated three first-order Medium Energy Grating (MEG) light curves:
(1) a \ion{Mg}{12} $\lambda\lambda 8.419;8.425$ light curve, (2) a
\ion{Mg}{11} $\lambda\lambda 9.169;9.228;9.231$ light curve, and
(3) a continuum light curve extracted from regions near the Mg lines that
are free of spectral lines. The exact wavelength ranges used to
create the light curves are 8.3962--8.4476 \AA\ for \ion{Mg}{12},
9.1457--9.1917 and 9.2052--9.2542 for \ion{Mg}{11}, and 8.0190--8.0674
and 8.4476--8.5060 for the continuum.

Figure~\ref{fig:1} shows light curves for \ion{Mg}{12}, \ion{Mg}{11},
and the line-free continuum as a function of the white dwarf spin
phase, using the optical ephemeris of \citet{hel1992}. The times
corresponding to the partial eclipses in the binary light curve
\citep{hoo2005} have been removed from these light
curves. Furthermore, the light curves are plotted in flux units to
facilitate direct comparison.  All three light curves show the
sinusoidal behavior noted in earlier studies
\citep[e.g.,][]{ros1988,all1998} and usually explained as due
to either the accretion curtain or occultation (see \S~1).

In addition to the broad modulation, the \ion{Mg}{11} light curve
shows two sharp spikes near spin phases $\phi = 1.0$ and $\phi = 1.2$
that have not been reported before. The spike at $\phi = 1.0$ is
significant at 2.9~$\sigma$ and the spike at $\phi = 1.2$ is
significant at 3.6~$\sigma$. Light curves for other He-like spectral
lines, most notably S, Si, and Ne, also show spikes at the same spin
phases but at lower ($<3~\sigma$) significance levels.

We note that none of the \ion{Mg}{12}, \ion{Mg}{11}, or continuum
light curves are centered around $\phi = 1.0$, the expected spin
maximum from the \cite{hel1992} optical ephemeris. This discrepancy
could indicate either a difference in the variations in optical and
X-ray emission \citep[see, e.g.,][]{bel2005} or that the ephemeris
needs to be updated. It is unclear from this dataset which explanation
is the more likely cause of the disagreement. In the remainder of the
letter, we assume that the maximum of the sinusoidal modulation
occurs when the  upper accretion column points away from the
observer, consistent with the predictions of both the accretion curtain and occultation
models.

This is not the first time that light curve information for individual
spectral lines has been used to determine physical properties of the
accretion column of a magnetic
CV. \citet{ter2001}, using {\it ASCA}\/ data for the polar V834 Cen,
showed that the high optical depth in the Fe K line resulted in
anisotropic radiation transfer through the accretion column,
collimating the Fe K emission. This collimation was detected as a
rotational modulation in V834~Cen Fe K light curve.

\section{Occultation Model}
To explain the spikes in the \ion{Mg}{11} light curve and the absence of
spikes in the \ion{Mg}{12} light curve requires that we consider the
specific conditions, chiefly the temperature ($T$) and density ($n$)
of the emitting plasma, under which a spectral line is produced. In
the remainder of this letter we assume that (1) the X-ray emitting
plasma is contained within the two accretion columns of \exhya, (2)
the two accretion columns point radially away from the white dwarf at
an angle $\alpha$ to the white dwarf spin axis, (3) the plasma is
heated to $T = 15.4$~keV \citep{fuj1997} by a stand-off shock at some
distance above the surface of the white dwarf, and subsequently cools
while settling onto the surface of the white dwarf, and (4) the
accretion columns are optically thin in the continuum and lines.

In the standard model for accretion onto a white dwarf, the
\citet{aiz1973} model, the hot, fully ionized, shocked plasma cools by
thermal Bremsstrahlung alone. However, other cooling processes, e.g.,
cooling by line emission or thermal conduction, also act on the plasma
and will result in deviations in the temperature profile from the
standard model. While the plasma cools and flows down towards the
white dwarf surface it recombines, producing a stratified ionization
state distribution with height. Given that spectral lines form under
specific physical conditions, i.e., temperature, density, and
ionization state, we can assume that each spectral line is formed at a
specific height, or a range of heights, above the white dwarf
surface. Combining this assumption with the fact that spectral lines
formed near the surface are more likely to be occulted by the white
dwarf as it rotates than lines that are formed high above the surface,
we naturally expect variations in the shape of spectral-line light
curves with height above the white dwarf.

To develop the occultation model, we begin with the simplifying
assumption that all emission originates from an infinitesimally small
point in space. To calculate the light curves produced by sources of
emission above the surface of the white dwarf, we assume that (a) the
white dwarf is represented by a sphere and is viewed at an inclination
$i$, where $i = 90\degr$ coincides with a viewing angle perpendicular
to the rotation axis of the white dwarf and (b) the emission is
located at a height $h$, expressed in white dwarf radii $R_{\rm WD}$, above the surface
at a latitude $b$ and longitude $l$. For this situation, the source of
emission will be visible to an observer at white dwarf spin phase
$\phi$ when the following condition is satisfied:
\begin{equation}
\label{eq:1}
\cos (l - 2\pi\phi) > -\frac{\tan b}{\tan i} -\frac{{\rm cot}\{{\arcsin}[1/(1+h)]\}}{(1+h)\cos b \sin i}.
\end{equation}

\begin{figure}
\plotone{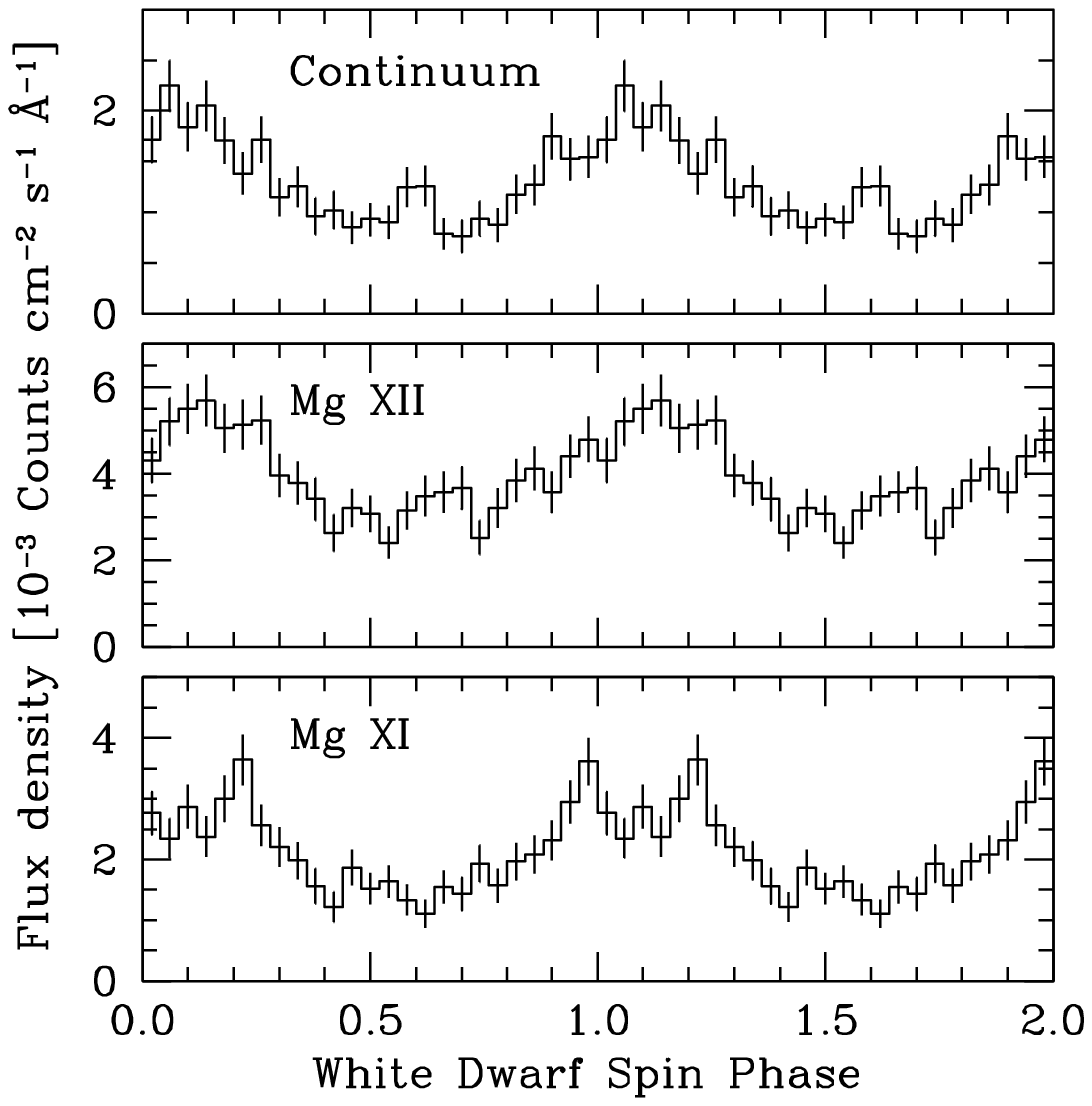}
\caption{MEG first order light curves for the continuum ({\it top})
and the \ion{Mg}{12} ({\it middle}), and \ion{Mg}{11} spectral
lines ({\it bottom}).}
\label{fig:1}
\end{figure}

In the remainder of the letter we use the inclination $i = 77.2\degr$
appropriate to EX~Hya \citep{beu2003,hoo2004}. We also assume that the
accretion columns are represented by cylinders that are perpendicular
to the white dwarf surface and that the diameters of the cylinders are
very small compared to the white dwarf radius. Furthermore, we
assume that the angle $\alpha$ represents the misalignment between the
white dwarf's rotation axis and the accretion region, i.e., $b =
\pm(90\degr - \alpha)$ represents the latitudes of the accretion
column and $\alpha = \delta + \epsilon$, where $\delta$ is the dipole
offset and $\epsilon$ is the the magnetic colatitude \citep[see
e.g.,][]{all1998}. We ignore the effects of absorption by the
accretion curtains, which produces the broad sinusoidal variation seen
in Figure~\ref{fig:1}.

With these assumptions, the emission from the lower accretion column
is occulted by the white dwarf for most of the time and is visible
only around $\phi \sim 0$, i.e., when the lower column points towards
the observer. The emission from the upper column behaves in the
opposite manner. There are two limiting cases: one where the emission
is on the surface ($h = 0$) and the other where the emission is
far above the surface ($h \gg 1$). In the former case, only
one of the emitting spots is visible at any time; the moment one spot
disappears from view the other appears. This will result in a flat
light curve. In the latter case, both spots of emission will be
visible at all times (except in the case of $i=\alpha=90\degr$), also
resulting in a flat light curve, but at twice the flux level as
compared to the $h=0$ case, since both columns are visible at all
times.

There are two critical heights in this model. First is the height
above the surface below which occultation occurs for the upper
accretion column: $h_{\rm top} = 1 / \sin(i+\alpha) - 1$. Emission at
$h \ge h_{\rm top}$ will not produce double-spiked light curves since
the upper accretion column is visible at all times. Second is the
height above the surface below which occultation occurs for the lower
accretion column: $h_{\rm bottom} = 1 / \sin(i-\alpha) - 1$. Using
similar arguments as above, emission at $h \ge h_{\rm bottom}$ will
produce completely flat light curves since $h_{\rm bottom} \ge h_{\rm
top}$. For emission at $h < h_{\rm top}$, two spikes will be visible
in the light curve. To illustrate the model, we show how the upper and
lower accretion columns are visible/occulted and how each contributes
to the total light curve.

\begin{figure}
\plotone{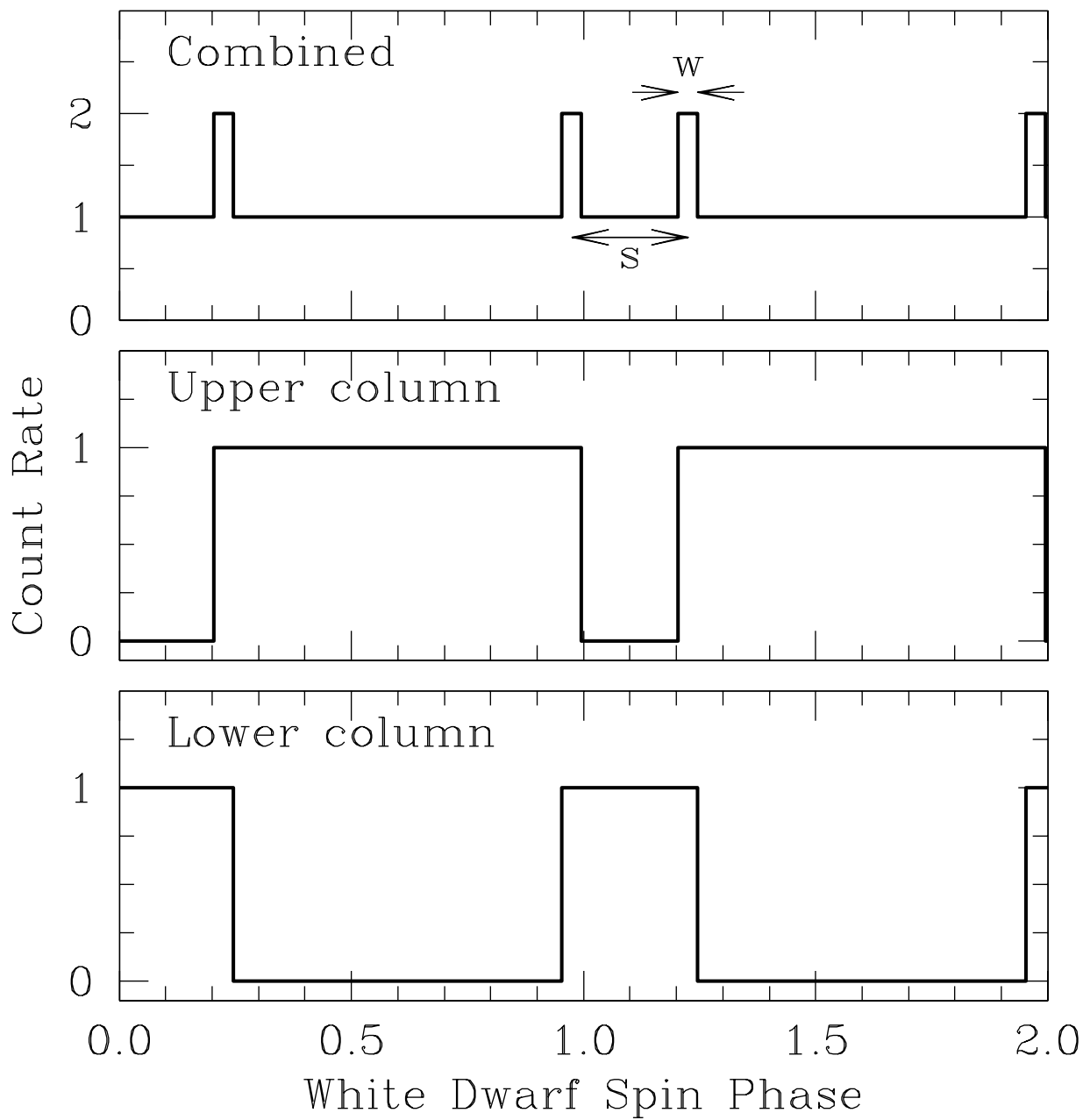}
\caption{Model light curves for $h = 0.0004$ and $\alpha = 18\degr$
for the lower accretion column ({\it bottom}), upper accretion column
({\it middle}), and combined light curve ({\it top}). The width $w$
and separation $s$ of the spikes are indicated in the top panel. The
light curves have been offset 0.1 in phase to account for the
difference between the optical and X-ray ephemeris (see text).}
\label{fig:2}
\end{figure}

Figure~\ref{fig:2} shows the light curve for two spots of emission:
one located in the upper accretion column at $l = 180\degr$, $b =
72\degr$, and $h = 0.0004$ and the other located in the lower accretion
column at $l = 0$, $b = -72\degr$, and $h = 0.0004$, i.e., $\alpha =
18\degr$ (these values correspond to the results presented later in
this letter). With this configuration, the upper accretion column
points away from the observer at $\phi = 0.0$, as predicted by both
the accretion curtain and occultation models. In Fig.~\ref{fig:2} we
have denoted as $w$ the width of one of the spikes (both spikes are
identical due to reasons of symmetry) and as $s$ the separation
between the two spikes. In the limit that $h$ tends to 0 we find that
$w = 0$ while in the limit that $h$ tends to $\infty$ we find that $w
= 0.5$, i.e., $w$ increases monotonically with $h$. The details of the
relationship between $s$ and $h$ depends on the inclination, but $s$
will also increase monotonically with $h$. 

Figure~3 shows contours of equal $w$ (black) and equal $s$ (gray) as a
function of $h$ and $\alpha$ for the inclination $i=77.2\degr$
of \exhya.  By measuring the width $w \la 0.04$ (this is an upper
limit since the whole spike fits in one phase bin) and separation
$s\sim0.25$ of the spikes in Fig.~\ref{fig:1}, we are able to
determine (1) the angle between the rotation axis and accretion
columns $\alpha \sim 18\degr$ and (2) the height of the emission above
the white dwarf surface $h \la 0.0004$ or $\la 4$~km for a white dwarf
radius of $10^9$~cm \citep{beu2003,hoo2004}. We can directly
compare $\alpha$ to $\delta + \epsilon$ (dipole offset plus magnetic
colatitude) from \citet{all1998}, who find that this angle can not be
larger than $23\degr$, consistent with our result. 
Furthermore, the absence of spikes in the \ion{Mg}{12} light curves
suggests that the \ion{Mg}{12} emission in at least one of the
accretion columns is not occulted by the body of the white
dwarf. Consequently, most of the \ion{Mg}{12} emission must originate
above $h_{\rm top} = 0.004 R_{\rm WD}$.

\section{Discussion and Conclusions}
In realistic accretion column models, the emission from a spectral
line does not originate from a point, but is distributed over a range
of heights in the column depending on the particulars of the density
and temperature profile. Hence, a spectral-line light curve can be
calculated by convolving the point source occultation model with the
line intensity vs.\ height profile, calculated from the temperature
$T(h)$ and density $n(h)$ profile in the accretion column as $I(h)
\propto \epsilon(T[h])\; n^2(h)\; dh$, where $\epsilon$ is the emissivity
of the line in  ${\rm photons~cm^{3}~s^{-1}}$.

The simplest model for $T(h)$ and $n(h)$ for radial accretion onto a
 white dwarf is presented by \citet{aiz1973}. The model assumes
that the shock height is small compared to the white dwarf radius so that
gravity can be assumed to be constant, and considers cooling only by
thermal Bremsstrahlung. The line intensity curves $I(h)$ for the Aizu
model are presented in Figure~4 of \citet{fuj1997}. They show that the
He-like ion line intensity curves peak below those of the H-like ions,
and that the H-like ion line intensity curves have long tails that extend
to large heights above the white dwarf surface. Both features tend to
produce H-like line light curves that are more featureless than the
He-like line light curves, but the difference is not sufficient to
explain the {\it Chandra\/} observations of the Mg line light curves:
assuming that the Aizu model describes the run of $T(h)$ and $n(h)$ in
the accretion column, we find that the occultation model predicts that
both He-like and H-like line light curves will show spikes. Given that
we found that the He-like \ion{Mg}{11} line light curve shows spikes,
but that the H-like \ion{Mg}{12} does not, we tentatively conclude that
the Aizu model does not produce the $T(h)$ and $n(h)$ profiles in the
accretion column of \exhya, and that modifications to the Aizu model must
be considered.

Several improvements to the Aizu model have been made in recent years.
\citet{wu1994} extended the model by including the effects of
cyclotron cooling, but these effects should be insignificant in \exhya\/
because of its weak magnetic field \citep[$\sim$0.2~MG;][]{eis2002}.
\citet{cro1999} included the effects of gravity on the accretion
column structure, finding lower shock temperatures and flatter
temperature profiles. Such a model may apply to \exhya, which other
evidence suggests may have tall shocks \citep{all1998}. If so, \exhya\
may have more hot material at large heights above the white dwarf
surface, which could increase the \ion{Mg}{12} $I(h)$ for large $h$
and thereby reduce the spikes in its light curve. \citet{wu2001}
calculated the effects on the accretion column structure due to
heating by the stellar surface and mass leakage at the base of the
column. Changes to the integrated X-ray spectra were neglible unless
the thickness of the mass leakage region approaches or exceeds 1\% of
the shock height, but large variations are produced in the $T(h)$ and
$n(h)$ profiles at heights where the Mg line emissivities nominally
peak. \citet{can2005} included the influence of the curvature of the
magnetic field on the $T(h)$ and $n(h)$ profiles, finding that
compressional heating due to the field geometry is as important as
radiative cooling and gravity in determinning the structure of the
post-shock flow. All the papers above employ the same closed
integral-form solution to solve the differential equations that
describe the accretion column structure. Other processes that
influence the $T(h)$ and $n(h)$ profiles such as electron-ion
equilibration \citep{sax2005} and thermal conduction \citep{spi1962}
can be included in hydrodynamical simulations. All these improvements
to the Aizu model change the temperature and density profiles and hence
change the spectral-line light curves, making it possible, provided
sufficient SNR, to use these light curves to constrain the run of
$T(h)$ and $n(h)$ and hence test the importance of various
processes in the model.

We presented, for the first time, white dwarf spin phase resolved
light curves for individual spectral lines for the bright IP
\exhya. 
The light curves for \ion{Mg}{11} show spikes never
seen before; two spikes around the maximum of the light curve. 
The width and separation of \ion{Mg}{11} light curve spikes allows for
a determination of (1) the angle between the rotation axis and the
accretion columns $\alpha = 18\degr$, (2) the height at which the
\ion{Mg}{11} emission formed, $h \sim 0.0004$, or $\sim 4$~km for a
white dwarf radius of $10^9$~cm. Since each spectral line is formed at
a different temperature and density and hence at a different height,
this type of light curve analysis can be used to map the distribution
of line emissivities, and hence the temperature and density structure,
along the accretion column of magnetic cataclysmic variables. Such a
detailed analysis of the accretion column of \exhya\ is feasible today
with {\it Chandra}, but requires a long dedicated ($\sim$500~ks)
observation to acquire sufficient photons to make high SNR, high
time-resolution light curves for many spectral lines spanning a wide
range of formation temperatures. The high spectral resolution of {\it
Chandra}\/ is critical, since it dramatically reduces (1) the
contribution of the continuum to the spectral-line light curves and
(2) the effects of blending, especially blending of lines formed at
different temperatures. Future X-ray missions, such as {\it
Constellation-X}, which have an order-of-magnitude larger collecting
area and similar or better spectral resolution compared to the HETG
instrument on {\it Chandra}\/, will be ideally suited to apply
spectral-line light curves techniques to a large sample of magnetic
CVs. An independent measurement of the temperature and density
structure of accretion column of magnetic CVs will be a valuable tool
in the study of shock physics and accretion onto compact objects in general.



\acknowledgments We thank H.~Tananbaum for the generous grant of
Director's Discretionary Time that made possible the {\it Chandra\/}
observations of EX~Hya. We acknowledge support from NASA through {\it
  Chandra} grant GO-4017X. NB was supported by NASA contract NAS8-39073
to the Chandra X-ray Center. CWM's contribution to this work was
performed under the auspices of the U.S.~Department of Energy by
University of California Lawrence Livermore National Laboratory under
contract No.~W-7405-Eng-48.

\begin{figure}
\plotone{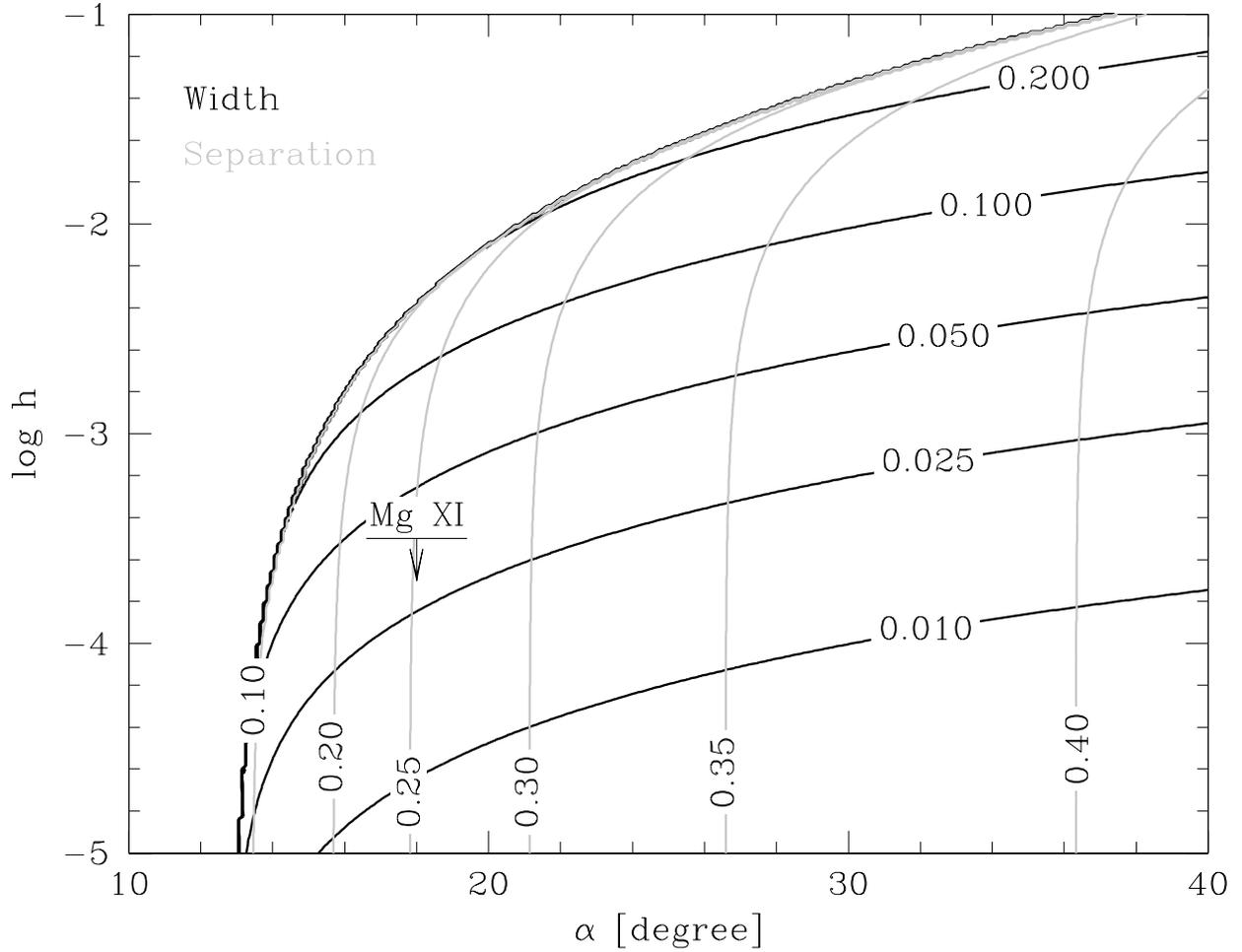}
\caption{Contours of equal width $w$ (black) and separation $s$ (gray)
for point source light curve for the inclination
$i=77.2\degr$ for \exhya. The approximate values for $w$ and $s$ from
the observed \ion{Mg}{11} light curve are indicated.}
\label{fig:3}
\end{figure}
%


\end{document}